\documentclass[fleqn,aps,pra,nofootinbib,groupedaddress]{revtex4-2}
\usepackage{graphicx}
\usepackage{amsmath}

\begin{document}

\title{\boldmath An analytic model for description of muonic oxygen X-ray time distribution in muonic experiments}

\author{P. Danev$^1$}
\email{Corresponding author: petar\_danev@abv.bg}
\author{I. Boradjiev$^{1,2}$}
\email{Corresponding author: borajiev@inrne.bas.bg}
\author{H.S. Tonchev$^1$}

\affiliation{$^1$ Institute for Nuclear Research and Nuclear Energy, Bulgarian Academy of Sciences, boul. Tsarigradsko ch. 72, Sofia 1784, Bulgaria} 
\affiliation{$^2$ University of Architecture, Civil Engineering and Geodesy, Boul. H. Smirnenski 1, Sofia 1046, Bulgaria}

 \date{\today}
 
 \begin{abstract}
We propose an analytical model and perform numerical simulations to study the time distribution of the characteristic muonic oxygen X-ray emission following muon transfer from muonic hydrogen to oxygen in a $H_2+O_2$ gas mixture. The model accounts for all fundamental processes that alter the  kinetic energy and spin distribution of muonic hydrogen atoms. The impact of the uncertainties in various experimental parameters on the precision 
of the computed results is studied in detail by means of Monte Carlo method. 
Verification against available experimental data reveals the potential of this approach for both description and parameter optimization in planning and analysis of muonic experiments.
 \end{abstract}

\maketitle
\section{Introduction}
\label{sec:intro}

The spectra of light atomic and molecular systems could be computed theoretically and measured experimentally with an unprecedented precision. Spectroscopy studies of such systems gave new information about the matter - antimatter duality \cite{Anderson2023} and led to the refinement of fundamental physical constants \cite{Schenkel2024, Alighanbari2020, Karr2023}. By replacing the electron in the hydrogen atom with a negatively charged muon, the relatively long-lived system of the exotic muonic hydrogen atom ($p\mu$) is formed. Due to the high mass of the muon in comparison with the electron, on average, $\mu$ is much closer to the proton and can better "feel" its structure. In 2010, the group of R. Pohl, by measuring the Lamb shift in $p\mu$, found the proton charge radius with very low uncertainty \cite{Pohl2010}. A large discrepancy between those results and previous experiments led to the so called "proton radius puzzle". As of now, this problem is generally considered solved in favor of the $p\mu$ spectroscopy data. 

The values of the charge and Zemach radii, characterizing the proton shape, can
be determined directly in a spectroscopic experiment.
The Zemach radius of the proton $R_Z$ is a fundamental quantity that combines the proton's charge distribution and magnetic dipole distribution. Though $R_Z$ has been been calculated in previous studies \cite{Dupays2003, Bakalov2005, Volotka2005, Friar2004}, its value remains insufficiently precise. 
The main factors limiting the precision of the value of Zemach radius are the uncertainties in the hyperfine splitting of the ground state of the muonic hydrogen, $\Delta E^{\text{hfs}}$, and the proton polarizability. A few experiments, including FAMU~\cite{Pizzolotto2020} and CREMA~\cite{Antognini2022}, are currently focused on the high precision measurements of $\Delta E^{\text{hfs}}$. 
These experiments leverage muon transfer in muonic hydrogen and its changes induced by resonant laser radiation \cite{Pizzolotto2020,  Antognini2022, Kanda2018}. 

The muonic spectroscopy experiments are complex, and achieving the required precision demands detailed understanding of all underlying processes. As for example, the muon transition from muonic hydrogen to oxygen and sulphur molecules, which has been investigated in \cite{Werthmuller1998, Mulhauser1993}. In these studies, by fitting to experimental data, a simple model expression for the collision-energy-dependent muon transition rate to oxygen has been obtained. However, recently, {Stoilov et al.}~\cite{Stoilov2023} derived a more precise expression for this quantity under specific assumptions, building on the results of the FAMU collaboration~\cite{Pizzolotto2021}. Furthermore, a comprehensive description of the transition of muons to heavier elements in gas mixture must also include the interaction of muonic hydrogen with the surrounding atoms and molecules. The relevant cross-sections and rates of elastic and inelastic scattering of $p\mu$ on hydrogen molecules, including the spin structure of $p\mu$, have already been provided by A. Adamczak in \cite{Adamczak2006}.  

An important stage in the planning and analysis of muonic experiments is the simulation of physical processes under realistic experimental conditions. These simulations are crucial for optimizing experimental parameters to observe a given phenomenon and maximize measurement effectiveness. 

In this study, we analytically model the time evolution of kinetic energy and spin distribution of muonic hydrogen in an $H_2+O_2$ gas mixture. This is achieved by incorporating the probabilities of the $p\mu$ atom transitioning between kinetic energy and spin states, undergo decay, or forming other exotic systems, into a transition matrix, which leads us to a system of ordinary differential equations of kinetic-rate type. Using the obtained muonic distribution, we compute muonic transfer rate from $p\mu$ to oxygen, and numerically investigate the time-dependent X-ray emission spectrum of muonic oxygen, which is directly observed in muonic experiments. 
Additionally, we calculate the transition rate of muons from hydrogen to oxygen nuclei in the laboratory reference frame under the assumptions for Maxwellian distribution for the low kinetic energies ($E \leq 0.1$ meV) of muonic hydrogen, and speculate on its high-energy functional behavior. To account for the uncertainties in the parameters and approximations in the model, we perform a thorough analysis by means of Monte Carlo method. Evaluation of the impact of these uncertainties on the precision of the simulation results is made
and those whose deviations contribute to the lowest-order uncertainties in the number of muon transitions to oxygen are identified. The applications of the proposed model is discussed, including its potential to optimize muonic experiments.
Similar models for computing the muon transfer rate have been used, for example, in FAMU~\cite{Mocchiutti2020} and CREMA~\cite{Nuber2023} collaborations' studies, however, our approach is independent of any specific experiment.

This paper is organized as follows. In section~\ref{sec:model}, we review  the analytical model. The model parameters and the approximations used  are discussed in section~\ref{sec:parameters}. In section~\ref{sec:Computatingtheuncertainties} we introduce estimators that quantify both the average behavior and the precision of our simulations. The main numerical results concerning the simulation precision are presented and discussed in section~\ref{sec:numericalResults}. A comparison with experimental data is provided in section~\ref{sec:comparisonWithExperiment}, and the conclusions  are outlined in section~\ref{sec:conlusion}.

\section{Muonic hydrogen decay model}\label{sec:model}

Modeling the time evolution of muonic hydrogen atoms in a gas mixture is a complex task due to the variety of processes that contribute to the change in the number and state of these atoms.
In this section, we present an analytical model to describe the time, energy, and spin distribution of the $p\mu$ atoms in a gaseous mixture of hydrogen and oxygen molecules. A key aspect of this investigation is determining the time evolution of the rate of muon transfer from hydrogen to oxygen in the gas. This quantity is essential for interpreting experimentally accumulated data \cite{Pizzolotto2020, Werthmuller1998}. 

In our model, the state of muonic hydrogen atom is described by two quantities – its kinetic energy and the total spin of the system. In the ground state of $p\mu$, the hydrogen nucleus and the muon spins can be coupled into singlet and triplet states, which will be denoted by $F_{\alpha} = 0, 1$ respectively. To simplify the problem, we discretize the kinetic energy into $n$ bins, denoted as $E_i$ for $i = 1, .., n$. The state of $p\mu$ is then represented as a $2n$-dimensional vector $N_{(i\alpha)}(t)=N(E_i,F_{\alpha},t)$, whose time evolution is governed by the relation: 

\begin{align}\label{eq:probability}
	N(t) = \text{e}^{L t} \, N(0),
\end{align}	
where the 
vector $N(0)$ represents the 
initial distribution of $p\mu$ atoms. The $2n\times 2n$ matrix $L$ encapsulates the contributions from various processes affecting the $p\mu$ state. For simplicity, $L$ is divided into two components: a diagonal matrix $L_d$ and a dense matrix $L_s$ such that $L=L_d+L_s$. 

The matrix $L_d$ accounts for the processes such as muon decay and nuclear capture rates, represented by $\lambda_0$, and the formation rates of muonic oxygen ($O\mu$), muonic deuterium ($d\mu$), and molecular ions ($pp\mu$) through the corresponding rates $\Lambda_{pO} =  \text{diag}\{\lambda_{pO}^0,\lambda_{pO}^1\}$, $\lambda_{d\mu}$ and $\lambda_{pp\mu}$. Explicitly, $L_d$ is given by 
\begin{align}
	L_d = &- (\lambda_0\, I_{2n} + \phi (c_p \lambda_{pp\mu} \, I_{2n} + c_d \lambda_{d\mu} \, I_{2n} + c_O\,\Lambda_{pO} )).  \label{eq:Ld} 
\end{align}
where, $I_{2n}$ is the $2n\times 2n$ identity matrix, $\phi$ is the number density of the atoms of the gas mixture in LHD units, and $c_p$, $c_d$, and $c_O$ are the number concentrations of hydrogen, deuterium and oxygen atoms, respectively. The $n$-dimensional vectors $\lambda_{pO}^{\alpha}\, (\alpha = 0,1)$ correspond to the energy-dependent muon transfer rate from hydrogen to oxygen. All transition rates except $\lambda_0$ are normalized to the liquid hydrogen density $N_{LHD}=4.25\, 10^{22}$ atoms/cm$^3$.

The transition matrix $L_s$ describes the elastic and inelastic scattering of muonic hydrogen by hydrogen molecules in the gas mixture:
\begin{align}\label{eq:Ls}
 L_s &= \phi c_p \begin{bmatrix}
		\lambda^{00,T}-\text{diag}\{{\cal I}_n^T\cdot\left(\lambda^{00} +\lambda^{01}\right)\} & \lambda^{10,T} \\
		\lambda^{01,T} & \lambda^{11,T}-\text{diag}\{{\cal I}_n^T\cdot\left(\lambda^{10} +\lambda^{11}\right)\}
	\end{bmatrix}
\end{align}
Here, ${\cal I}_n$ is an $n$-dimensional column vector of ones, and $\lambda^{\alpha \beta} (\ \alpha, \beta=0,1)$ are $n\times n$ matrices comprising $p\mu$ scattering rates on hydrogen. Specifically, $\lambda^{\alpha \beta}_{ij}$ denotes the transition rate of a $p\mu$ atom from a quantum state with spin $F=\alpha$ and kinetic energy $E_i$ to a state with $F=\beta$ and $E_j$.

The primary quantity of interest is the transfer rate from hydrogen to oxygen in the gas, expressed as the number of transfers from hydrogen to oxygen per unit time: 
\begin{align}
    \frac{\text{d}N_{p\mu \rightarrow O\mu}}{\text{d}t}(t)=\phi c_O \,
    {\cal I}_{2n}^T\cdot \Lambda_{pO}\cdot N(t)  .
    \label{muonTransferRate}
\end{align}
The emitted characteristic X-rays following the muon transfer to oxygen occur on a timescale much shorter ($\sim 10^{-13}$s \cite{Measday2001}) than other simulated processes ($\sim 10^{-6} - 10^{-10}$s). Therefore, we can approximate the experimentally measurable X-ray emission rate as directly proportional to the muon transfer rate to oxygen in the gas. 
In the subsequent discussion, we will consider these two rates as equivalent:
\begin{align}
   \text{d}N_O/\text{d}t\equiv \text{d}N_{p\mu \rightarrow O\mu}/\text{d}t.
\end{align}

\section{Simulation parameters and sources of errors}\label{sec:parameters}

\subsection{Physical constants}\label{sec:physicalconstants}

In our computations, we use the most recent values of the physical constants necessary to calculate the muon decay rate according to the presented model. These are provided in Table~\ref{tab:physicalconstants}.

\begin{center}
	\begin{table}[h]
		\caption{\label{tab:physicalconstants} Physical constants and their uncertainties used in the simulations
		}
	\vspace{5pt}
		\small\rm
		\centering
		\begin{tabular}{| c | c | c | c |}
			\hline
			Quantity & Label & Value & Ref. \\ [0.5ex] 
			\hline
			Total muon decay rate & $\lambda_{0}$ & $(4.66501\pm 0.00014)\times 10^5$ s$^{-1}$ & \cite{Mocchiutti2018}\\
           $pp\mu$ formation rate & $\lambda_{pp\mu}$& $(2.01\pm 0.07)\times 10^{6}\ s^{-1}$ 
			 & \cite{MuCap2015} \\
          $d\mu$ formation rate & $\lambda_{d\mu}$ & $(1.64\pm0.16) \times 10^{10}\ s^{-1}$ & \cite{Mocchiutti2020}\\
          Muonic hydrogen mass  & m$_{p\mu}$ & $(1043.927647\pm 0.000023)\times 10^6$ eV/c$^2$ & \cite{CODATA2018} \\
          Oxygen molecule mass & m$_{O_2}$ & $(29806\pm 2)\times 10^6$ eV/c$^2$& \cite{CODATA2018}\\
			\hline
		\end{tabular}
	\end{table}
\end{center}

\subsection{Experimentally controllable quantities}

Some of the simulation parameters depend on the specific conditions of a particular experiment. In order to maintain a certain level of concreteness, we focus on the physical variable values used in the measurements reported by {Werthm\"uller et al.} \cite{Werthmuller1998}. These values are summarized in Table~\ref{tab:Controlablequantities}. By using parameter values corresponding to a real experiment, we can compare the predictions of our model with the experimentally observed data provided in \cite{Werthmuller1998}. 
\begin{center}
	\begin{table}[h]
		\caption{\label{tab:Controlablequantities} Experimental parameter values used in the simulations
		}
	\vspace{5pt}
		\small\rm
		\centering
		\begin{tabular}{| c | c | c | c |}
			\hline
			Quantity & Label & Value & Ref \\ [0.5ex] 
			\hline
			Oxygen atoms concentration & $c_O$ & $(39.60\pm 0.40)\times 10^{-4}$   & \cite{Werthmuller1998}\\
           Deuterium nuclei concentration & $c_d$& $(1.50\pm0.05)(1-c_O)\times10^{-4}$ & \cite{Werthmuller1998} \\
          Temperature & $T$ & $(297.15\pm 0.50)$ K & \cite{Werthmuller1998}\\
          Pressure  & $P$ & $(15.06\pm 0.05) $ bar & \cite{Werthmuller1998} \\
			\hline
		\end{tabular}
	\end{table}
\end{center}


\subsection{Number density of atoms $\phi$}

To calculate the number density of atoms normalized to LHD, $\phi$ in Eq.~\ref{eq:Ld}, we have adopted the ideal gas approximation. This simplifies $\phi$ to a function of the ratio of the pressure and temperature: $\phi(P/T)=P/(k_B T N_0)\approx 0.3408\times P/T$, where $P$ is in units of bars. Since the hydrogen constitutes the predominant portion of the gas mixture, and at the temperature of interest only a small portion of the gas particles engage in inelastic scattering processes, this approximation is sufficiently accurate for our study. For the set of the experimental parameters used in our simulations, deviations due to this approximation remain within 1\% \cite{HydrogenTables2018}. 

In our analysis, the uncertainties in the number of muon events stemming from $\phi$ are effectively accounted for through the uncertainties in temperature and pressure.

\subsection{Initial distribution of muonic  hydrogen}\label{sec:initialDistribution}

It is known, that the initial kinetic energy distribution of the muonic hydrogen, $N(0)$, does not conform to a simple Maxwellian profile. Experimental and theoretical studies suggest the existence of a high-energy $p\mu$ population \cite{Werthmuller1998, Markushin1994}. Therefore, for $p\mu$ atoms in both singlet and triplet spin states, we adopt the "Two Component Model", proposed in \cite{Werthmuller1998}, where a fraction $\kappa$ of $p\mu$ atoms resides at an initial energy $E_0^{\text{high}}$ of approximately $20$ eV and the remaining fraction (1-$\kappa$) follows Maxwell statistics. In our simulations, we use $\kappa=0.4 \pm 0.1$, which approximates the results presented in \cite{Markushin1994} for the corresponding experimental parameters. 
We also assume that, at the onset of the process, the muonic hydrogen atoms are statistically distributed between the two hyperfine spin states: $1/4$ in the singlet state ($F=0$) and $3/4$ in the triplet state ($F=1$). 

To account for the uncertainty in the position of the high-energy component in the initial $p\mu$ distribution, we introduce an uncertainty in $E_0^{\text{high}}$ as $E_0^{\text{high}}=\bar{E}_0^{\text{high}}\pm \Delta E_0^{\text{high}}$, where $\bar{E}_0^{\text{high}}=20$ eV and $\Delta E_0^{\text{high}}=2$eV, approximately matching the width of an energy bin.

\subsection{Numerical accuracy and time discretization}

We verify that in computing the muon transfer rate, the numerical errors are several orders of magnitude smaller than the uncertainties introduced by the approximations and experimental parameter uncertainties discussed earlier. Consequently, these numerical errors can be considered negligible in our analysis.

We compute the time evolution of the muonic hydrogen distribution using discrete time steps. Since the matrix $L$ governing this evolution is time-independent, this discretization does not introduce any errors. Therefore, the errors in the muon transfer rate arise solely from the accumulation of the numerical error during the evolutionary steps. Given the high precision of the used numerical methods, the numerical errors inherent in these approaches are minimal and do not significantly impact the overall accuracy and precision of the results. 

\subsection{$\mu H$ atoms kinetic energy discretization}

From Eqs.~(\ref{eq:probability}) and (\ref{muonTransferRate}), it is evident that the errors due to the muon hydrogen kinetic energy discretization can be effectively incorporated into the uncertainties of the energy-dependent quantities - muonic hydrogen scattering rates, transfer rate of muons to oxygen, and initial distribution. When these discretization errors are smaller than the respective uncertainties, as is the case in the regime considered here, they do not significantly affect the precision of the simulation. 
 
\subsection{Muonic hydrogen scattering rates $\lambda^{\alpha,\beta}$} 

The elastic and inelastic scattering rates of muonic hydrogen, $\lambda_{ij}^{\alpha,\beta} \, (\alpha,\beta=0,1)$, at a temperature of $T=300$ K, used in our simulations,  
are taken from Adamczak~\cite{Adamczak2006}. These rates are 
averaged over the Boltzmann distribution of the initial rotational 
levels of the $H_2$ molecule and over the Maxwellian kinetic energies of these molecules, for the given gas temperature. 
Furthermore, downwards spin-flip transition $F=1 \rightarrow F=0$  is given for a single averaged value of the projection $F_z$ of the initial total spin $F=1$.
$\lambda^{\alpha,\beta}$ are computed with a relative uncertainty of $\sigma^{\lambda^{\alpha,\beta}_{ij}}/\mu^{\lambda^{\alpha,\beta}_{ij}}\lesssim 10^{-5}$ and an absolute uncertainty $\sigma^{\lambda^{\alpha,\beta}_{ij}}\lesssim 10^{5}s^{-1}$ up to $100$eV, ensuring that their uncertainty can be safely neglected.

\subsection{Transfer rate of muons from $p\mu$ to oxygen $\lambda_{pO}(E)$}
\label{lambda_pO_unc}

The collision-energy-dependent transfer rate of muons from hydrogen in singlet state to oxygen $\lambda_{pO}^{0s}(E^c)$ is a key quantity for studying the time evolution of muon transfer to oxygen. Only recently was $\lambda_{pO}^{0s}(E^c)$ obtained with sufficient accuracy \cite{Stoilov2023}, thus enabling to conduct more precise simulations of such physical processes.  

To obtain the $p\mu$ kinetic energy-dependent transfer rate with respect to the laboratory reference frame, $\lambda_{pO}^0(E,T)$, we proceed as follows. 
First, we express the collision energy in terms of the solid angle $\Omega$ between the momenta of the $p\mu$ atom and the (point) $O_2$ molecule, and their kinetic energies 
$E$ and $E_{O_2}$ correspondingly.
Then, we average $\lambda_{pO}^{0s}(E^c(E,E_{O_2},\Omega))$ over the kinetic energy of oxygen molecule $E_{O_2}$ and the solid angle $\Omega$, weighted with the Maxwell distribution $f_{M}(T,E)$ for temperature $T$,
\begin{align}
	\lambda_{pO}^0(E,T) = \int_0^{\infty}\int_{\Omega} f_{M}(T,E^c) 
	\lambda_{pO}^{0s}(E^c)\frac{\text{d}\Omega}{4\pi}\text{d}E_{O_2}. \label{lambdapO}
\end{align}

The use of Maxwell distibution of $O_2$ atoms in the gas is just an approximation, since the gas is a mixture of real gases. The associated uncertainty is assumed to be incorporated into the uncertainty of $\lambda_{pO}^{0s}(E^c)$ (obtained under the assumption of Maxwell distribution over $E^c$). Moreover, during the first tens of nanoseconds until termalization occurs, the $O_2$ distribution is time-dependent; however, in the subsequent discussions this time dependence will be neglected. The uncertainty of $\lambda_{pO}^{0s}(E^c)$ is propagated into the uncertainty of $\lambda_{pO}(E,T)$ through Eq.~(\ref{lambdapO}).

The energy dependence of the muon transfer rate $\lambda_{pO} (E)$ used in our simulations is characterized by relatively small uncertainties for $p\mu$ energies up to $E\simeq 0.1$eV. For higher energies ($E \gtrsim 0.1$eV), however, the values of $\lambda_{pO} (E)$ are known with lower precision \cite{Stoilov2023}. To address this, we consider two scenarios: a moderate, where a smooth extrapolation of $\lambda_{pO} (E)$ uncertainty is applied, and a conservative scenario, where for $E>0.1$eV the uncertainty is assumed to be significantly larger. The $95 \%$ confidence intervals for both cases are shown in the upper panels of figure~\ref{fig:lambdapOcomp}.

We further assume that the muon transfer rate from the triplet state ($F=1$) of muonic hydrogen to oxygen, ($\lambda^1_{pO}(E)$), is equal to the transfer rate from the singlet state, i.e. $\lambda^1_{pO}(E)=\lambda^0_{pO}(E) \equiv \lambda_{pO}(E)$. This approximation may introduce minor deviations from the actual time dependence of the muon transfer rate only in the first few tens of nanoseconds, prior the thermalization of the gas. However, its impact is negligible at later times, as the $p\mu$ atoms in the excited state rapidly converge to their spin ground state ($F=0$).

\section{Estimators: Sample Mean, Sample Standard Deviation, Sample Relative Standard Deviation} \label{sec:Computatingtheuncertainties}

The time interval of $1\mu$s is divided into $1000$ time slices of size $\Delta t_{j+1} = t_{j+1}-t_j=1$ns, where $j=0,1,...1000$. 
For each parameter $X$ presented in section~\ref{sec:parameters}, and at each time $t_j$, we  conducted a large number of Monte Carlo simulation runs ($N_{\text{run}}=1000$), 
evaluating the respective muon transfer rates to oxygen, $(\text{d}N_O/\text{d}t(x_i;t_j))$, with $1\leq i\leq N_{\text{run}}$. 
For each run, a random value $x_i$ of the parameter $X$ was selected, based on the distribution of its uncertainty, while all other quantities were kept at their mean values.
Finally, for the $(j+1)^{\text{th}}$ time bin, the relative standard deviation ($RSD$) of $(\text{d}N_O/\text{d}t(x_i;t_j))$ due to uncertainties in $X$ is computed as:
\begin{align}  
RSD^{\text{d}N_O/\text{d}t}(X;t_j)=\frac{\sigma^{\text{d}N_O/\text{d}t}(X;t_j)}{\mu^{\text{d}N_O/\text{d}t}(X;t_j)},
\end{align}
where the mean and standard deviation of the muon transfer rate are defined as
\begin{align}
&\mu^{\text{d}N_O/\text{d}t}(X;t_j)=\frac{1}{N_{\text{run}}} \sum_{i=1}^{N_{\text{run}}}\frac{\text{d}N_O}{\text{d}t}(x_i;t_j),\\
&\sigma^{\text{d}N_O/\text{d}t}(X;t_j)=\sqrt{\frac{1}{N_{\text{run}}-1} \sum_{i=1}^{N_{\text{run}}}\left(\frac{\text{d}N_O}{\text{d}t}(x_i;t_j)-\mu^{\text{d}N_O/\text{d}t}(X;t_j)\right)^2}. \nonumber
\end{align}
Since we are primarily interested in the impact of the parameter $X$'s uncertainty on the time distribution of the muon transfer rate from hydrogen to oxygen $\text{d}N_O/\text{d}t$, in the following text, we adopt a simplified notation
$RSD^{\text{d}N_O/\text{d}t}(X;t_j)\equiv RSD(X)$  
and $\mu^{\text{d}N_O/\text{d}t}(X;t_j) \equiv \mu(X)$. 
If necessary, the time variable $t_j$ will be explicitly indicated. 

\begin{center}
	\begin{figure}[h]
    \centering
			\includegraphics[width=16pc]{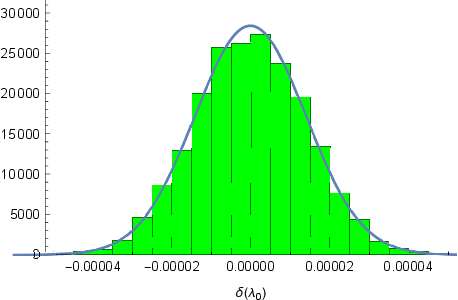}\qquad
			\includegraphics[width=16pc]{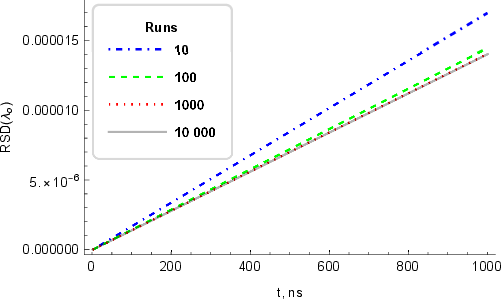}
			\caption{(Left)  
				Distribution of the relative deviation of the muon transfer rate to oxygen,  $\delta(\lambda_{0i}) = [\text{d}N^{\lambda_0}_O/\text{d}t (\lambda_{0i}) - \text{d}N^{\lambda_0}_O/\text{d}t(\mu(\lambda_{0}))] / [\text{d}N^{\lambda_0}_O/\text{d}t(\mu(\lambda_{0}))]$, at $t=1000$ns, caused by the uncertainty in $\lambda_0$, for $N_{\text{run}}=10 000$. The solid line represents the corresponding normal distribution. 
				(Right) Relative standard deviation of $\text{d}N_O/\text{d}t$ resulting from the uncertainties in $\lambda_0$, calculated for a selected number of simulation runs.  }\label{fig:lambda0Sigma}
	\end{figure}
\end{center}

Two remarks are in order. First, our simulations show that relatively small uncertainties in the normally distributed parameters lead to uncertainties in the muon transfer events $\text{d}N_O/\text{d}t(t)$ that follow a distribution very close to normal. This is illustrated in figure~\ref{fig:lambda0Sigma}(Left), which presents a histogram of the relative uncertainty in $\text{d}N_O/\text{d}t(t={1000}$ ns), caused by the uncertainty in $\lambda_0$, for 10 000 runs. The solid curve represents a normal distribution with the same mean and standard deviation. 

Second, to demonstrate that the chosen number of simulation runs is optimal for our investigation, we performed simulations with different values of $N_{\text{run}}$ and plotted the results in figure~\ref{fig:lambda0Sigma}(Right). The curves corresponding to $N_{\text{run}}=1000$ and $N_{\text{run}}=10000$ are visually indistinguishable.

\section{Propagation of uncertainty in the muon transfer rate. $RSD$ dynamics. Analysis of model precision.}\label{sec:numericalResults}

In this section, we present the impact of the uncertainties in the physical variables on the precision of the simulated muon transfer rate to oxygen $\text{d}N_O/\text{d}t$. 
We consider a time interval of $1\mu$s since this is approximately the time span of the most informative experimental observation.

\subsection{Uncertainties in the physical constants} \label{sec:physicalconstantsEffect}
\begin{center}
	\begin{figure}[h]
    \centering
			\includegraphics[width=16pc]{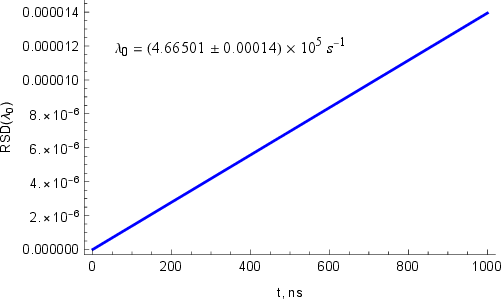}
          \quad   \includegraphics[width=16pc]{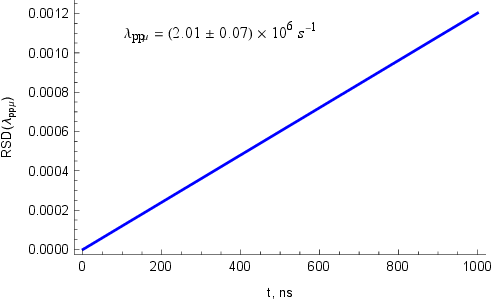}\\
    	\quad\includegraphics[width=16pc]{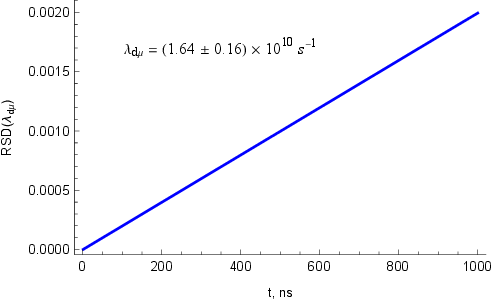}\qquad
		\includegraphics[width=15pc]
  {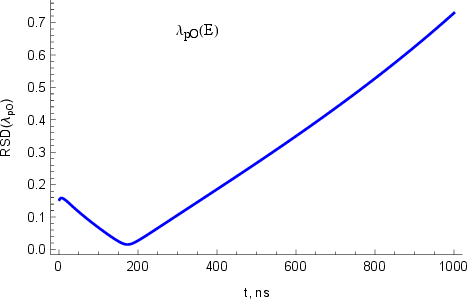}\\
  \includegraphics[width=16pc]{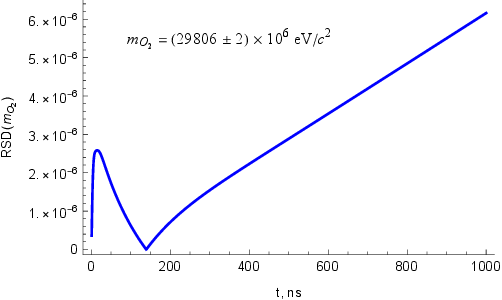} \		 \includegraphics[width=16pc]{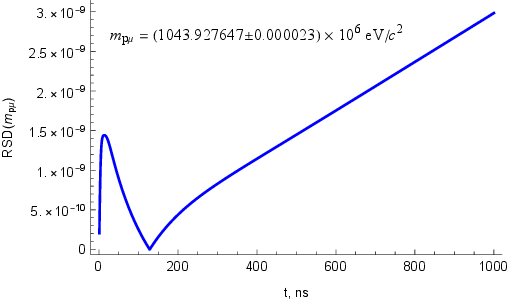}\caption{Relative standard deviation of muon transfer rate determined by uncertainties in various physical constants used in the simulations, plotted as a function of time.}\label{fig:physicalconstantsEffect}
	\end{figure}
\end{center}
The impact of the uncertainties of the physical constants used in the simulations on the time distribution of the relative standard deviation of muon transfer rate is shown in figure~\ref{fig:physicalconstantsEffect}. The $RSD$ due to the quantities listed in Table~\ref{tab:physicalconstants} range from less than 1\% (for $\lambda_{pp\mu}$, $\lambda_{d\mu}$) to significantly smaller values for other constants.

However, the relative 1$\sigma$ deviations in $RSD(\lambda_{pO})$ caused by the uncertainty of $\lambda_{pO}(E)$ for a moderate scenario (see section \ref{lambda_pO_unc}) 
are non-negligible. For times shortly after the formation of $p\mu$ atoms ($0\div 150$ ns), these deviations reach approximately $15\%$ and by $t=1000$ ns they exceed $70\%$. 

Given the value of $\text{d}N_O/\text{d}t(t)$ decreases exponentially over time, the dominant influence of the uncertainty in $\lambda_{pO}(E)$ on the muon transfer rate is observed at the beginning of the process, specifically for times $t\lesssim 100$ns, as seen in figure~\ref{fig:lambdapOcomp}.

\subsection{Uncertainties in the controllable experimental parameters}\label{sec:parametersEffect}

\begin{center}
	\begin{figure}[h]
    \centering
			\includegraphics[width=16pc]{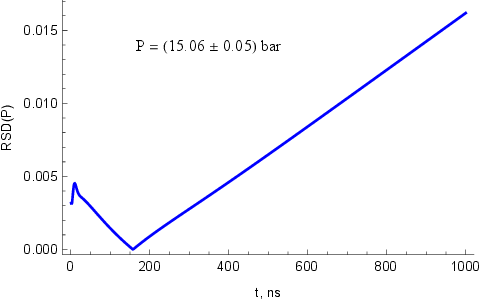}\quad  \includegraphics[width=16pc]{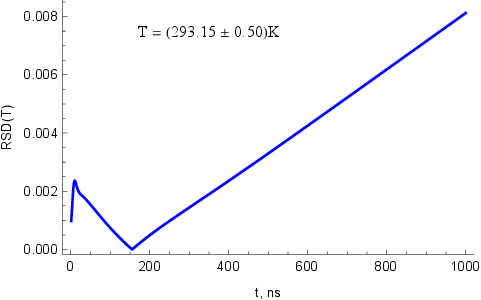}\\
             \includegraphics[width=16pc]{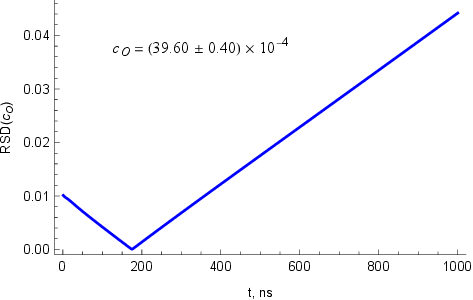}\quad  \includegraphics[width=16pc]{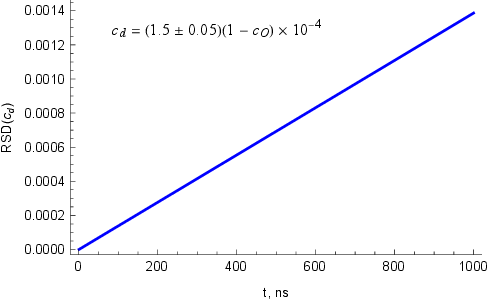}
			\caption{Relative standard deviation of muon transfer rate determined by uncertainties in the experimentally controlled parameters used in the simulations, plotted as a function of time.}\label{fig:parametersEffect}
	\end{figure}
\end{center}

Unlike the physical constants, the other parameters used in the simulations depend highly on the conditions of the modeled experiment. In this study, we utilized the mean values and the uncertainties of the controllable parameters corresponding to the experiment described in \cite{Werthmuller1998} (given in Table~\ref{tab:Controlablequantities}) to verify the results of our simulations. As shown in figure~\ref{fig:parametersEffect}, 
the $RSD$ of the muon transfer events due to inaccuracies in the experimentally controlled parameters $P$, $T$, $c_O$, and $c_d$ is significantly larger than that resulting from the uncertainties in the physical constants. As discussed in section~\ref{sec:scaling} the effect of the uncertainties in these parameters on the experimentally observed data in precise muonic experiments can be estimated.


\subsection{Uncertainty in the initial condition} \label{sec:parametersEffect}

\begin{center}
	\begin{figure}[h]
    \centering
			\includegraphics[width=16pc]{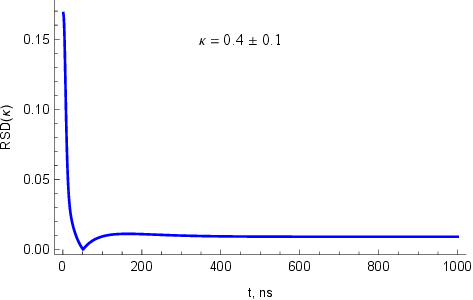}\quad
			\includegraphics[width=16pc]{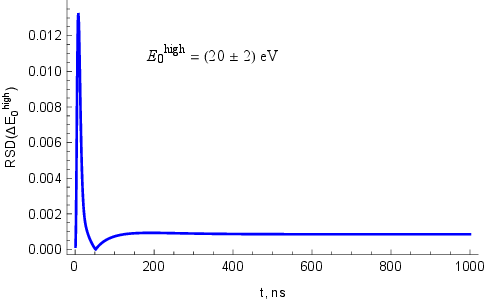}
			\caption{Relative standard deviation of muon transfer time distribution, determined by uncertainties in the initial energy distribution between the two energy components (left panel) and energy of the high-energy component (right panel) of the muonic hydrogen atoms. 
		} \label{fig:initialState}
	\end{figure}
\end{center}

The relative standard deviation in the muon transfer rate to oxygen, arising from the assumptions regarding the initial state of $p\mu$ atoms as discussed in section~\ref{sec:initialDistribution}, is shown in figure~\ref{fig:initialState}. The uncertainties in the fraction of the high-energy component $\kappa$ and its mean energy ${\bar{E}^{\text{high}}}$ (both with normal distribution and standard deviations respectively $\sigma{^\kappa}$ and $\sigma^{\bar{E}^{\text{high}}}$), result in significant variations in $\text{d}N_O/\text{d}t$ at the first moments after the formation of $p\mu$ atoms. As expected, the initial energy distribution has a major impact during the thermalization phase of $p\mu$ atoms, while its influence diminishes significantly - by approximately a factor of $20$ - once termalization is complete. The figure demonstrates that the $RSD$ due to uncertainties in the high-energy component fraction is an order of magnitude larger than that caused by its mean energy for the chosen uncertainties.

\subsection{Uncertainty in $\lambda_{pO} (E)$}

\begin{center}
	\begin{figure}[h!]
    \centering
			\includegraphics[width=17pc]{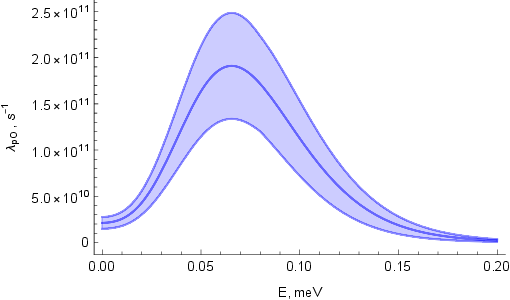}\quad
			\includegraphics[width=17pc]{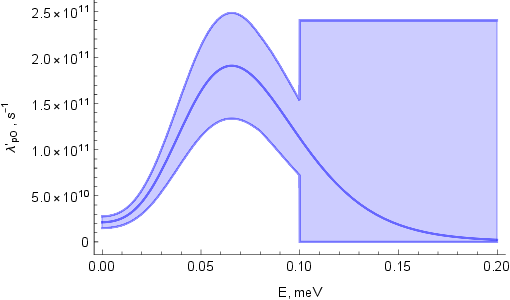}\\
			\includegraphics[width=17pc]{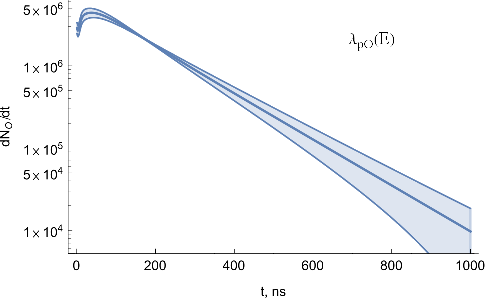}\quad
			\includegraphics[width=17pc]{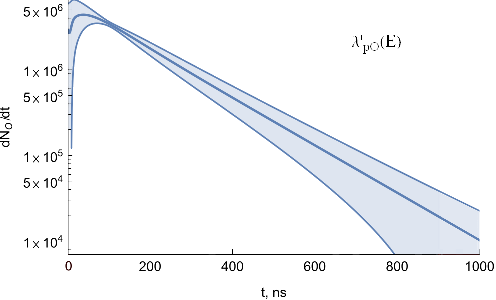}\\
			\includegraphics[width=17pc]{lambdapO_sigmarel_1000.eps}\quad
			\includegraphics[width=17pc]{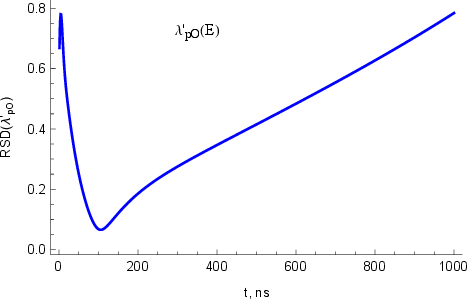}
			\caption{The energy dependence of the muon transfer rate $\lambda_{pO} (E)$, along  with its $95\%$ confidence interval, is displayed in the top panels. For $E\le 0.1$eV, the confidence interval is taken from \cite{Stoilov2023}, while for $E>0.1$ eV, two scenarios are considered: moderate (left panel) and conservative (right panel). The corresponding muon transfer rate $\text{d}N_O/\text{d}t$ and relative standard deviation $RSD(\lambda_{pO})$ computed with the respective $\lambda_{pO} (E)$ are presented in the middle and bottom rows. The muon transfer rate is plotted on a logarithmic scale.} \label{fig:lambdapOcomp}
	\end{figure}
\end{center}

Simulations are performed under both moderate and conservative uncertainty scenarios for $\lambda_{pO} (E)$, as depicted in the upper pannels of figure~\ref{fig:lambdapOcomp}. The resulting muon transfer rates are presented in the middle panels of of the same figure, with shaded areas representing the standard deviation of $\text{d}N_O/\text{d}t$. 

The relative standard deviations due to the uncertainties in $\lambda_{pO} (E)$ as functions of time are shown in the bottom panels of figure~\ref{fig:lambdapOcomp}. A comparison of the figures indicates that increasing the uncertainty in $\lambda_{pO} (E)$ for $E>0.1$ eV has minimal impact on the muon transfer rate for 
times $t\gtrsim 100$ ns. However, during the initial stages of the transfer process, the impact is significant - the relative standard deviation for the two scenarios differs by a factor of five.

\subsection{Analytical considerations}\label{sec:analyticalConsiderations}

The uncertainty propagation in the muon transfer rate, characterized by its relative standard deviation, can be approximated as
\begin{align}
	RSD(\{X\}) &\approx \frac{1}{\mu(\{X\})} \sqrt{ \sum_{k=1}^q \left( \sigma^{X_k} \frac{\partial}{\partial X_k} \frac{\text{d}N_O}{\text{d}t}(\{\mu^{X}\})\right)^2},  \label{RSD_uncertaintyPropagation}\\
	\{X\} &= X_1,X_2...,X_q, \, \{\mu^{X}\} = \mu^{X_1}, \mu^{X_2},...\mu^{X_q}, \notag
\end{align}
where $\sigma^{X_k}$ is the standard deviation of $X_k,\,k=1,2,...q$, and the derivative with respect to $X_k$ is evaluated at the mean values $\mu^{X_k}$ of the respective parameters. 
The derivation of Eq.~(\ref{RSD_uncertaintyPropagation}) assumes that approximating $\text{d}N_O/\text{d}t(\{X\})$ by a first-order expansion is valid, and the random variables $X_k$ are statistically independent. 

The parameters $c_O$, $\Lambda_{pO}$, and $\phi(P/T)$ appear as multiplicative factors of $N(t)$ in the expression (\ref{muonTransferRate}) for the muon transfer rate from hydrogen to oxygen. This results in a distinct time of minimum uncertainty, $t_0$, which differs from the initial time of the process, particularly for the uncertainties associated with $c_O$, $\Lambda_{pO}$, $P$, and $T$  (figures~\ref{fig:physicalconstantsEffect}, \ref{fig:parametersEffect}).  

Approximate values of the time of minimum uncertainty $t_0$ in $RSD(X)$, for single parameters $X=c_0, P, T,\Lambda_{pO}$, can be found as a solution of the following equation:
\begin{align}
\frac{\text{d}N_O}{\text{d}t}(\mu^X + \sigma^X;t_0) - \mu(X;t_0) = 0 .
\label{eq:t0}
\end{align}
In the general case, the solution of Eq.~(\ref{eq:t0}) for $t_0$, and the slopes of $RSD(X)$ at $t_0$, lack simple closed-form analytical expressions. For that reason, we refrain from presenting overly simplified formulas including linearization of the matrix exponents, though for some specific parameter combinations such may be attainable.  


Parameters such as $m_{O_2}$ and $m_{p\mu}$ (figure~\ref{fig:physicalconstantsEffect}), whose uncertainties are incorporated in the aforementioned ones are not further considered in this context. Additionally, the impact of uncertainties in the initial conditions on the scaling with parameter uncertainties and the time dependence of $RSD(\kappa)$ and $RSD(\Delta E_0^{\text{high}})$ (figure~\ref{fig:initialState}) is not extensively discussed here, as these aspects can be analyzed in a similar manner.

To elucidate some key characteristics of the time-dependent behavior of $RSD(X)$, in the following,
we provide specific examples and additional commentary on the $RSD(\{X\})$ time dependence and its scaling with parameter uncertainties.

\subsection{Scaling of $RSD$ with parameter uncertainties
}\label{sec:scaling}

\begin{center}
	\begin{figure}[ht]
    \centering
			\includegraphics[width=16pc]{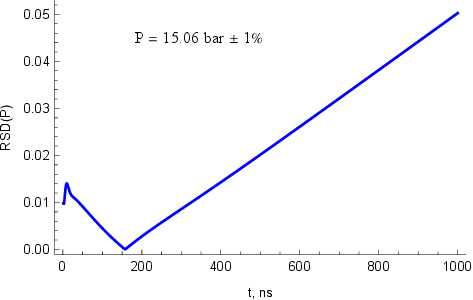}\quad
			\includegraphics[width=16pc]{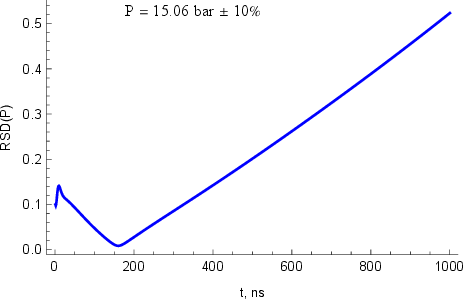}
			\caption{Relative standard deviation of the muon transfer rate 
				due to pressure uncertainty, shown for $\sigma_P/P=1$\% (Left) and $\sigma_P/P=10$\% (Right), as a function of time. }\label{fig:PressureSigma}
	\end{figure}
\end{center}

\begin{center}
	\begin{figure}[ht]
    \centering
			\includegraphics[width=16pc]{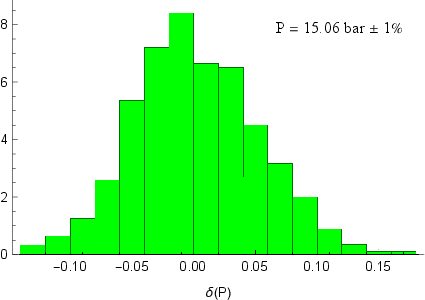}\quad
			\includegraphics[width=16pc]{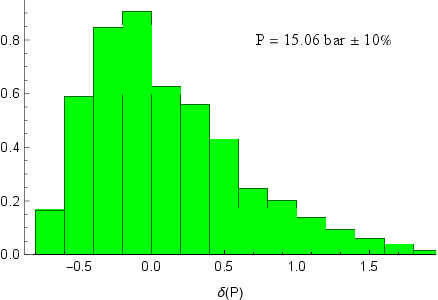}
			\caption{Distributions of the relative deviation of the muon transfer rate to oxygen,  $\delta(P_i) = [\text{d}N^{P}/\text{d}t (P_i) - \text{d}N^{P}/\text{d}t(\mu(P_{0}))] / [\text{d}N^{P}/\text{d}t(\mu(P_{0}))]$, at $t=1000$ns, due to the uncertainty in $P$, for $N_{\text{run}}=1000$. 
			 Histograms for relative standard deviations $\sigma_P / P=1$\% (left panel) and $\sigma_P / P =10$\% (right panel) are shown.
} \label{fig:PressureHistograms} 
	\end{figure}
\end{center}

Expression (\ref{RSD_uncertaintyPropagation}) indicates that for small deviations of the experimental parameters, the resulting change in $\text{d}N_O/\text{d}t$ must scale nearly linearly with the change of the parameter standard deviation. An example is shown in figure~\ref{fig:PressureSigma}, where the $RSD(P)$ is plotted for pressure deviations of 1\% (left panel) and 10\% (right panel), respectively. Comparing these plots reveals that the relative standard deviation increases approximately ten times. Furthermore, in this example, we simulated a 10\% uncertainty in the pressure - a value significantly overestimated for modern precision experiments. For such large uncertainties, the distribution of muon transfer rates at $t=1000$ns deviates from the normal one, as illustrated in figure~\ref{fig:PressureHistograms}. For smaller deviations in the experimental parameters, the assumption of linear scaling holds even more accurately.

\subsection{Remarks on the $RSD$ time dependence}

\begin{center}
	\begin{figure}[h]
  \begin{center}
	\includegraphics[width=20pc]{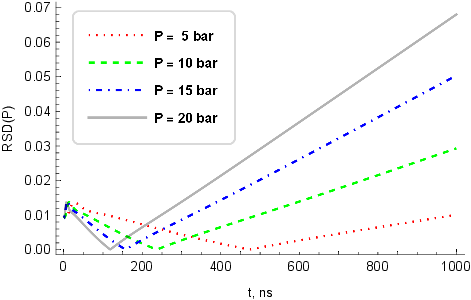}
		\caption{Relative standard deviation of the muon transfer rate due to uncertainties in  pressure for four different mean values of its distribution. The dotted red, dashed green, dot-dashed blue and solid gray lines correspond to mean values of $P=5, 10, 15$ and 20bar, respectively. In all cases the pressure uncertainty is~1\%. }\label{fig:Pressurecomparison}
   \end{center}
	\end{figure}
\end{center}

In the presented simulations, figures~\ref{fig:physicalconstantsEffect}, \ref{fig:parametersEffect}, \ref{fig:lambdapOcomp}, \ref{fig:PressureSigma}, \ref{fig:Pressurecomparison},
we observed that, in general, the time behavior of the relative standard deviation defined by various parameters tends to increase at large times. As we have already pointed out, the rate of change of $\text{d}N_O/\text{d}t$ in time is linear in first order of approximation.
However, since the mean value of the muon transfer rate $\text{d}N_O/\text{d}t$ decreases exponentially with time, the effect of parameter uncertainties on the standard deviation becomes smaller at later times. 

As we have already mentioned, for a given combination of experimental parameters, there exists a time interval where the impact of uncertainties in some of the parameters on the relative standard deviation becomes minimal. In the time dependence of $RSD$ due to uncertainties in $P, T, 
c_O, \lambda_{pO}$, a time of minimum uncertainty, $t_0$, is observed at approximately a few hundred nanoseconds, for the range of change of the mean values of parameters we are interested in. This phenomenon is attributed to faster (slower) $p\mu$ atoms depletion in the gas when the particular parameter has larger (smaller) value.

For instance, this effect is illustrated for a few pressure values in figure~\ref{fig:Pressurecomparison}. As the pressure $P$ increases from 5bar to 20bar, the time corresponding to the lowest $RSD$ shifts from $t_0\approx 500$ns to $t_0\approx 150$ns. A similar trend is observed for other parameters, reflecting the underlying relationship between these parameters and the muon transfer rate $\text{d}N_O/\text{d}t$, as described by Eqs.~(\ref{eq:probability}, \ref{eq:Ld}).

This effect has several potential applications. For instance, identifying the time interval of low $RSD$ allows for improved comparison of results between experiments and/or numerical simulations. For a specific experiment, it can also assist in calibration processes. Moreover, by optimizing the experimental parameters to exploit this effect, it may be possible to enhance the precision of the measurements, potentially by reducing the impact of parameter uncertainties during the critical interval of observation.

\section{Simulations and comparison with experimental data}\label{sec:comparisonWithExperiment}

\begin{center}
	\begin{figure}[h!]
    \centering
			\includegraphics[width=16pc]{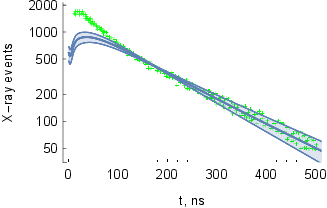}\quad
			\includegraphics[width=16pc]{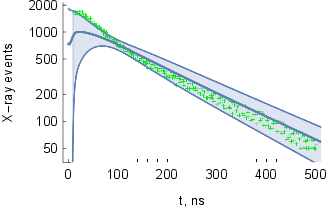}
			\caption{Simulations of the X-ray emission event rate (solid purple line) alongside  measured data from \cite{Werthmuller1998} (green "+" symbols). 
			The solid purple line represents the mean of all computed curves $\text{d}N_O/\text{d}t$ on a logarithmic scale, while the shadowed area denotes the $1\sigma$ standard deviation due to parameter uncertainties. The left and right panels correspond to the moderate and conservative scenario for the energy-dependent muon transfer rate to oxygen,   $\lambda_{pO}$ and $\lambda'_{pO}$ as shown in figure~\ref{fig:lambdapOcomp}, respectively. }\label{fig:WerthmullerComparison}
	\end{figure}
\end{center}

In order to test the precision and accuracy of our model versus a real-world scenario, we simulated the muon transfer rate from hydrogen to oxygen using the experimental parameters from \cite{Werthmuller1998} as specified in section~\ref{sec:parameters}.
In line with the analysis in section~\ref{sec:numericalResults}, we account for the uncertainties in the parameters that have the most significant impact on the simulations precision - namely, the pressure $P$ and temperature $T$ of the gas mixture, the oxygen concentration $c_O$, the initial distribution of $\mu p$, and the transfer rate of muons to oxygen, $\lambda_{pO}$. 
The simulation results showing the transfer rates from hydrogen to oxygen as a function of time along with their corresponding standard deviation are plotted alongside the experimental data from Werthm\"uller et al. \cite{Werthmuller1998} in figure~\ref{fig:WerthmullerComparison}. 
 
The simulations incorporate the transfer rate $\lambda_{pO}(E)$ from \cite{Stoilov2023}, using both moderate and conservative scenarios for its uncertainty, as shown in figure~\ref{fig:lambdapOcomp}. The results for these scenarios are displayed in the left and right panel of figure~\ref{fig:WerthmullerComparison}, respectively. The measured X-ray event rates are expected to be proportional to the muon transfer rate on a logarithmic scale, with the scaling factor depending on specific experimental factors, such as detector efficiencies, the number of observed characteristic frequencies, the initial population of $p\mu$ atoms in the target, and more. This scaling factor is determined by matching the average value of $\text{d}N_O/\text{d}t$ from the simulation to the experimental data fit at $t=170$ns for $\lambda_{pO}$ and at $110$ns for $\lambda'_{pO}$, as these times correspond to periods where the relative standard uncertainty in the muon transfer event rate is approximately minimal.

Figure~\ref{fig:WerthmullerComparison} shows a very good agreement between the simulation and the experiment at times $t \gtrsim 150$ns. 
However, a notable discrepancy in the slope of the emission event rate is observed for $t\lesssim 100$ns. This discrepancy results in a slight upward shift of the simulated curve  relative to the experimental data for $t \gtrsim 150$ns in figure~\ref{fig:WerthmullerComparison} (right)), as the matching is performed at time when the experimental behavior deviates from the expectations. 
This deviation is challenging to explain solely through the uncertainties in the physical constants (Table~\ref{tab:physicalconstants}) or experimental parameters (Table~\ref{tab:Controlablequantities}). 

Several factors may contribute to this discrepancy. For instance, our approximation of the initial $p\mu$ distribution as a "two-component" model may be too simplistic. However, it could partly explain the discrepancy as the uncertainties stemming from this approximation could be significant in the first tens of nanoseconds as can be seen in the left panel of figure~\ref{fig:initialState}. 
A second potential reason is the lack of a reliable expression for the energy-dependent muon transfer rate, $\lambda_{pO}(E)$, for energies $E$ higher than $0.1$eV. 
In this case, we have to assume that the uncertainty in $\lambda_{pO}(E)$ is larger for $E>0.1$ eV. When the conservative uncertainty estimate is used, all experimental data points fall within or very close to the $1\sigma$ uncertainty band, as shown in the right panel of figure~\ref{fig:WerthmullerComparison}. 

The observations suggest that the energy-dependent muon transfer rate $\lambda_{pO}(E)$ for $E>0.1$eV could be higher than currently assumed. This may be attributed to effects arising from the internal structure of $O_2$ and $p\mu$. Investigation of this hypothesis would require a dedicated study. In principle, the value of $\lambda_{pO}(E)$ could be tested through an experiment designed to measure the muon transfer rate as a function of the collision energy,  with high resolution, during the first tens of nanoseconds after $p\mu$ formation, before thermalization occurs.

\section{Conclusion}\label{sec:conlusion}

We study the characteristic time-dependent X-ray emission resulting from muon transitions to the lower energy states in oxygen after their transfer from muonic hydrogen. In the developed model, the kinetic energy and spin distribution of $p\mu$ atoms is evolved in time by accounting for the possible channels that alter their state. The decay and scattering of muonic hydrogen, as well as its probability of transferring to another constituent of the surrounding medium, are modeled by using the relevant transition rate matrix. The impact of the uncertainties in the experimental parameters is investigated by means of Monte Carlo method, and the main sources of uncertainty are identified. The results obtained from the proposed model are in a good agreement with the available experimental data, especially at relatively long times. However, within a first hundred nanoseconds, there is a discrepancy that may be due to an unaccounted effect or inaccuracies in the rates used. Thus, the model provides a realistic description of processes involving muonic hydrogen, while different sources of uncertainty are assessed.
Moreover, the presence of a minimum in the relative standard deviation can serve as a benchmark for comparing experiments and numerical simulations, aid in calibration processes, and enhance measurement precision by minimizing the impact of parameter uncertainties during the observation period.
These results could be useful and may find application in the planning and analysis of both current and future muonic experiments.

\acknowledgments

This work was supported by the Bulgarian Science Fund under contract
KP-06-N58/5 / 19.11.2021.
The authors would like to express their gratitude to A. Adamczak for providing the muonic hydrogen scattering rates.




\end{document}